\documentclass[fleqn,usenatbib]{mnras}
\DeclareSymbolFont{cmletters}{OML}{cmm}{m}{it}
\DeclareMathSymbol{v}{\mathalpha}{cmletters}{"76}

\usepackage{ae,aecompl}
\usepackage{times}
\usepackage{soul}

\usepackage{tabularx,ragged2e,booktabs,caption}
\usepackage{amsmath}
\usepackage{amssymb}
\usepackage{epsfig}
\usepackage{graphicx}
\usepackage{ifthen}
\usepackage{latexsym}
\usepackage{rotating}
\usepackage{times,epsf}
\usepackage{txfonts}
\usepackage{varioref}
\usepackage{verbatim}
\usepackage{array}
\usepackage{url}
\usepackage{color}
\usepackage[T1]{fontenc}
\usepackage{epstopdf}
\usepackage{ulem}

\def\gsim{\mathrel{\raise.5ex\hbox{$>$}\mkern-14mu
             \lower0.6ex\hbox{$\,\sim$}}}
\def\lsim{\mathrel{\raise.3ex\hbox{$<$}\mkern-14mu
             \lower0.6ex\hbox{$\,\sim$}}}

\newcommand{\be}{\begin{equation}}
\newcommand{\ee}{\end{equation}}
\newcommand{\bea}{\begin{eqnarray}}
\newcommand{\eea}{\end{eqnarray}}

\title{A ring-of-fire in the pulsar magnetosphere}
\author[I. Contopoulos, P. Stefanou]
       {I. Contopoulos$^{1,2}$\thanks{E-mail: icontop@academyofathens.gr}, P. Stefanou$^3$\\
$^1$ Research Center for Astronomy and Applied Mathematics, Academy of Athens, Athens 11527, Greece\\
$^2$ National Research Nuclear University (MEPhI), Moscow 115409, Russia\\
$^3$ Section of Astrophysics, Astronomy and Mechanics, Department of Physics, University of Athens, Athens 15783, Greece
}

\begin{document}

\maketitle

\label{firstpage}

\begin{abstract}
\citet{C19} proposed that a dissipation zone develops in the magnetosphere of young pulsars at the edge of the closed-line region beyond the light cylinder. This is necessary in order to supply the charge carriers that will establish current closure through the equatorial and separatrix current-sheets. In the present work, we propose to investigate in greater detail this region with a simplified model that we would like to call the {\it `ring-of-fire'}. According to this simple model, the dissipation zone is a narrow reconnection layer where electrons and positrons are accelerated inwards and outwards respectively along Speiser orbits that are deflected in the azimuthal direction by the pulsar rotation. After they exit the reconnection layer, the accelerated positrons form the positively charged equatorial current-sheet, and the accelerated electrons form the negatively charged separatrix current-sheet along the boundary of the closed-line region. During their acceleration, particles lose only a small part of their energy to radiation. Most of their energy is lost outside the dissipation region, in the equatorial and separatrix current sheets. Our simple model allows us to obtain high-energy spectra and efficiencies. The radiation emitted by the positrons in the equatorial current-sheet forms a very-high energy tail that extends up to the TeV range. 
\end{abstract}


\section{A simplified model}

The pulsar magnetosphere is the large scale electromagnetic system that transports Poynting energy from the central rotating magnetized neutron star to dissipative regions at large distances. This global picture has been observationally confirmed by comparing the neutron star spindown energy loss and the bolometric power of the pulsar wind nebula. In several cases, the two are found to be correlated \citep[e.g.][]{ZZF18}. Most recently, X-ray and $\gamma$-ray observations suggest that in more than 250 pulsars, a non-negligible fraction of the total spindown energy loss is also channelled into particles that emit pulsed high-energy radiation in the vicinity of the central neutron star \citep[e.g.][]{VBY11}.

Over the last decade, a lot of effort has been invested in understanding the origin of particle acceleration and dissipation in the pulsar magnetosphere. \citet{C07a,C07b} suggested that we need to move beyond the ideal force-free  magnetospheric solution of \citet{CKF99} and introduce dissipation. \citet{KKHC12, LST12,G13} proposed to quantify magnetospheric dissipation via various ad-hoc non force-free global parametric prescriptions of finite conductivity. High energy light curves and spectra were obtained by injecting particles at arbitrary positions with arbitrary Lorentz factors. \citet{KHK14,KHKB17} pursued this approach further and argued that, in order to conform with observations, magnetospheric dissipation must take place mostly outside the light cylinder and around the equatorial current-sheet. 
\begin{figure}
 \centering
 \includegraphics[width=8cm,height=6cm,angle=0.0]{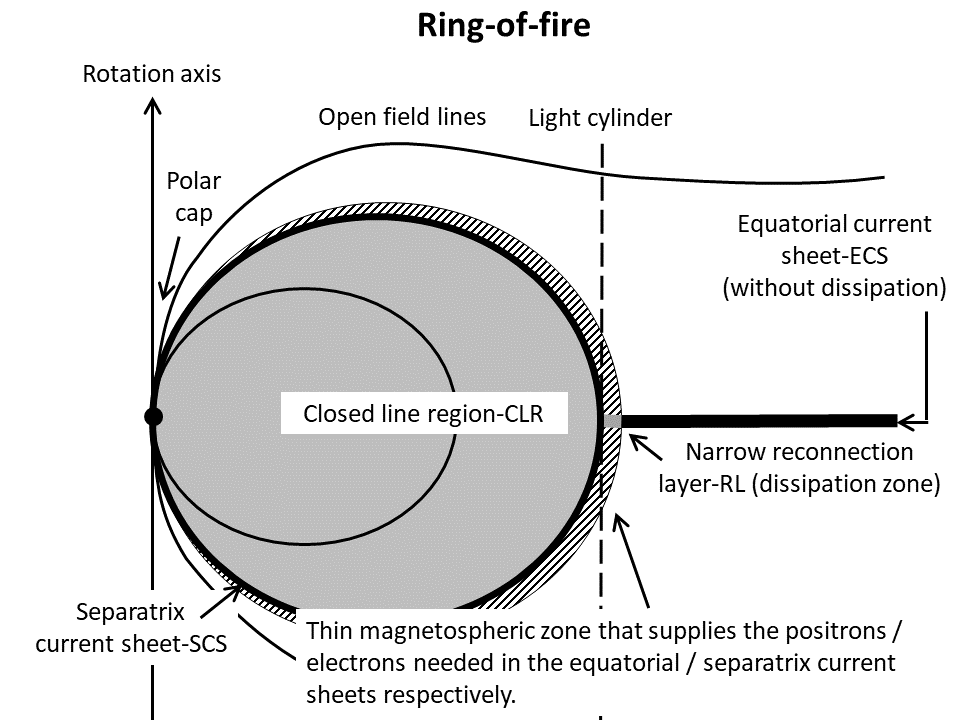} 
 \includegraphics[width=8cm,height=6cm,angle=0.0]{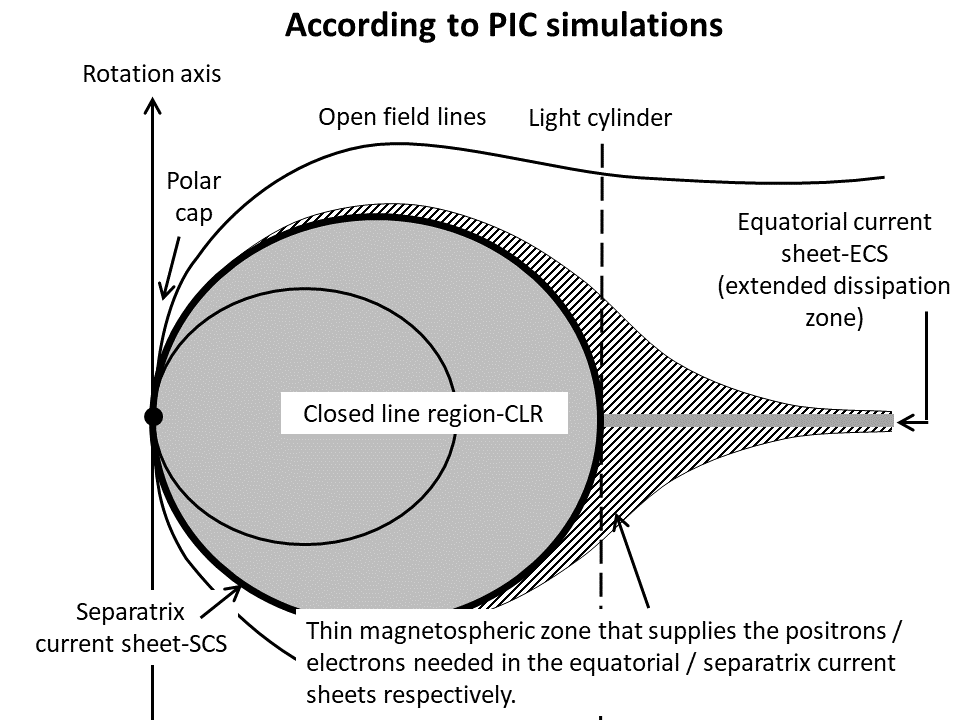}
\caption{Schematic of the magnetosphere (not drawn to scale). Shown: neutron star, closed and open field lines, separatrix (thick solid line), thin magnetospheric zone that supplies the charges of the current sheets (dashed region), equatorial current sheet and dissipation zone. Upper plot: Simplified model of the ring-of-fire according to Paper~I. Width of dissipation zone $\Delta\ll r_{\rm lc}$ (see text for details). Lower plot: Magnetospheric configuration resulting from PIC numerical simulations. The dissipation zone is extended, but most of the dissipation takes place near the light cylinder.}
\label{picture}
\end{figure}

Meanwhile, the astrophysical community decided that it is time to investigate the problem `ab-initio' with global Particle-in-Cell (hereafter PIC) simulations, hoping that these simulations would objectively pinpoint the origin of magnetospheric dissipation \citep[e.g.][]{PS14, Cetal16a, Cetal16b,Ketal18}. Despite their impressive animations of accelerated particle trajectories, we believe it is too early for global PIC simulations. The latest state-of-the-art simulations show high magnetospheric dissipation that cannot be safely discerned from the numerical one (compare for example Fig.~1c of \citet{TSL13} and Fig.~6 of \citet{CPPS15}, with the final conclusion of \citet{PS14}). Moreover, in order to relate the numerical results to observations, one needs to increase the Lorentz factors and magnetic field values used in the simulations by several orders of magnitude via some arbitrary scaling prescription \citep{PS18,Ketal18}. Global PIC simulations may yield important hints about the origin of the high-energy radiation in pulsars, we nevertheless deem their results still inconclusive.

A much simpler approach was proposed in \citet{C07a,C07b,CKK14} and more recently in \citet[][hereafter Paper~I]{C19}, in which the pulsar magnetosphere is considered to be everywhere ideal and force-free {\it except} in finite dissipative magnetospheric regions. In other words, magnetic field lines are the infinitely conducting `rotating wires' of an electric circuit that transfers electromagnetic energy from the central `generator' to specific dissipation loads at large distances. Such approach may some day lead to hybrid ideal force-free--PIC simulations in which the PIC computational effort will be focused only in the regions of interest where particle acceleration and high-energy radiation take place, and not everywhere in the magnetosphere where particles simply follow magnetic field lines. 

According to the model proposed in Paper~I, dissipation occurs right beyond the tip of the closed-line region (hereafter CLR; also known as the `dead zone') on the light cylinder. This is where the positively charged equatorial return-current-sheet connects to the negatively charged separatrix return-current-sheet along the boundary of the CLR. The dissipation zone develops precisely at that position in order to supply the electrons and positrons that are needed to support the electric charge and electric currents of the magnetospheric current-sheets (equatorial and separatrix), and thus establish global electric current closure. It is easy to see that, if $\kappa$ is the multiplicity of pair formation in the electrostatic gap above the polar cap, a {\it thin magnetospheric zone} of width $\delta\approx r_{\rm pc}/(2\kappa)\ll r_{\rm pc}$  inside the rim of the polar cap contains all the charges needed to support the magnetospheric current-sheets ($r_{\rm pc}$ is the radius of the polar cap around the magnetic axis). These charges are supplied to the current-sheets at the other end of this thin zone, in the equatorial dissipation zone beyond the light cylinder. In Paper~I, we assumed that the dissipation zone is narrow, with width $\Delta\approx r_{\rm lc}/\kappa$, where 
$r_{\rm lc}\equiv c/\Omega$ is the radius of the light cylinder, and $\Omega$ is the angular velocity of stellar rotation. In young pulsars $\kappa\gg 1$, and therefore, $\Delta\ll r_{\rm lc}$. This is why we decided to name the pulsar dissipation region the {\it `ring-of-fire'}\footnote{The name alludes to the rim of the Pacific Ocean where many earthquakes and volcanic eruptions occur.}. 

As we will see in the discussion below, this is a simplification. In particular, the equatorial current-sheet is spacelike \citep[fig.~6 of][]{T06}, and therefore, the supply of its charge carriers must be gradual and cannot be limited to a very narrow region (Kalapotharakos and Kazanas, private discussion). Indeed, PIC numerical simulations suggest that the dissipation zone extends some distance much larger than $\Delta$ along the equatorial current sheet \citep[e.g.][]{CPPS15}. Nevertheless, the total potential drop accross the dissipation zone remains the same as that accross the thin magnetospheric zone. We also expect that the largest fraction of the dissipation and the resulting particle acceleration still take place within a distance of about $\Delta$ from the light cylinder where poloidal magnetic field lines enter the dissipation zone almost perpendicularly. Beyond that distance, we expect only a small amount of extra dissipation and particle acceleration, both distributed gradually over a radial distance much larger than $\Delta$ (see figure~\ref{picture}). It is our understanding that present-day computational power does not allow us to study numerically global pulsar magnetospheres with pair-multiplicities $\kappa\gg 1$ \citep{C16}. This is why, in this series of papers, we opted for a semi-analytic approach in the framework of the simple model of a narrow equatorial dissipation region, and dissipationless separatrix and equatorial current sheets.

Without loss of generality, in what follows we consider only aligned rotators ($B$ along $\Omega$ at the poles), in which case electrons are accelerated inwards and positrons outwards. In the context of the present paper, magnetospheric dissipation means transfer of electromagnetic (Poynting) energy to particles, and {\it not} particle energy loss to radiation. We will see in the next section that the accelerated particles lose only a small part of their energy inside the reconnection layer, and most of it outside, therefore, {\it the high-energy radiation zone is much more extended than the dissipation zone}. This is consistent with the recent numerical results of \citet[][their fig.~18]{Ketal18}. Particles are accelerated in the radial direction by the same electric field as the one in the ideal force-free magnetosphere just above and below the reconnection layer, namely
\begin{eqnarray}
&&E=E_r=-B_z\left(\frac{r}{r_{\rm lc}}\right)\ .
\end{eqnarray}
Notice that, for a reconnection layer that lies outside the light cylinder, $|E_r|> |B_z|$.

\begin{figure*}
 \centering
 \includegraphics[width=18cm,height=9cm,angle=0.0]{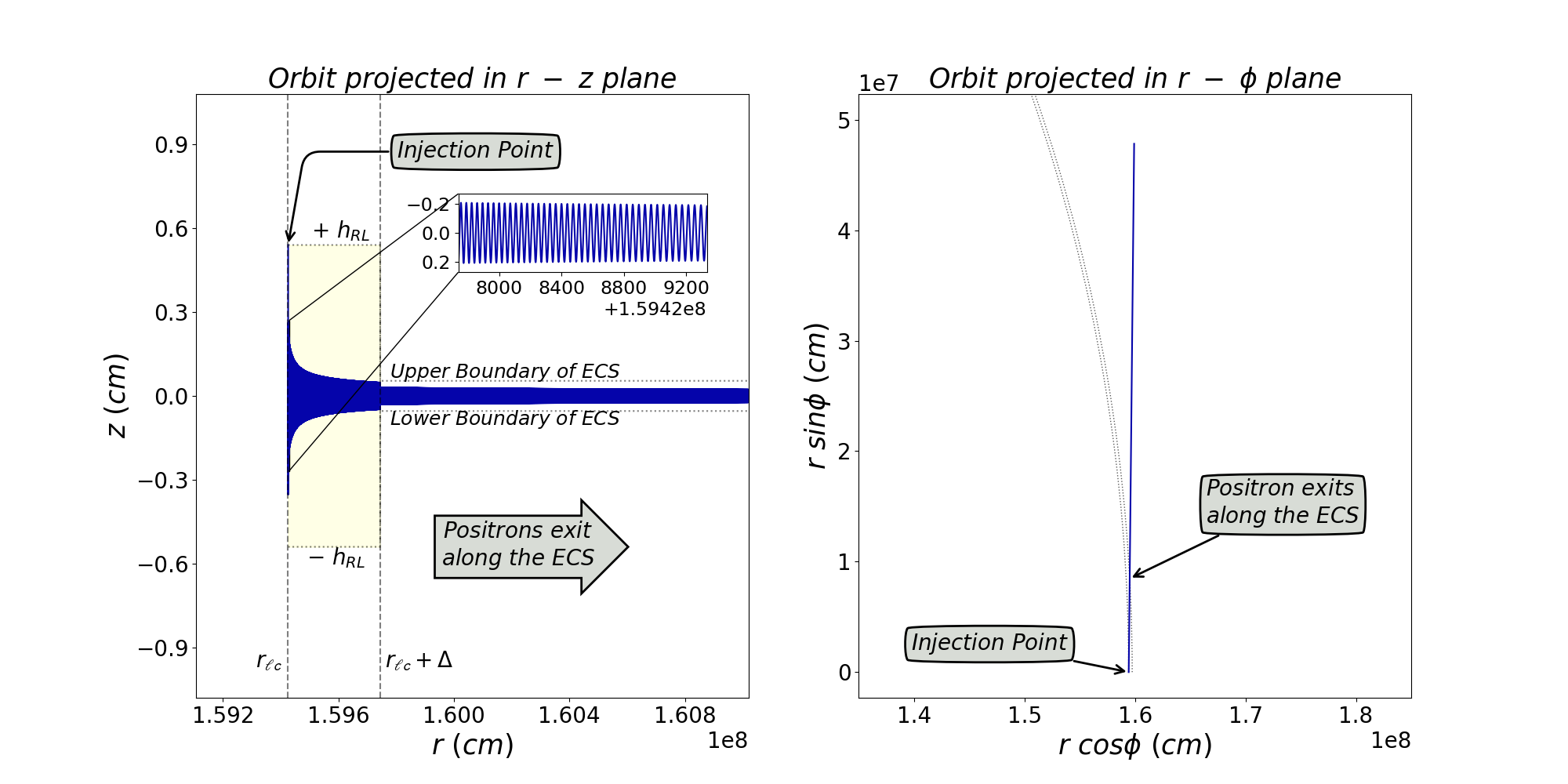}
\caption{3D trajectories of positrons injected at the inner edge of the equatorial reconnection layer with $\Gamma_{\rm inj}=500$ (blue line). Shown also various characteristic positions of the orbit. Left plot: projection of 3D orbit in the $(r,z)$ plane with detail of the Speiser orbit. Notice the smallness of the $z$-scale. Right plot: projection of 3D orbit in the equatorial $(r,\phi)$ plane (reconnection layer shown from above). The positrons are accelerated outwards by the radial electric field, and are deflected in the $\phi$ direction. After they exit the reconnection layer, they enter and support the positively charged dissipationless equatorial current-sheet (ECS) where they experience no further acceleration, and radiate away their energy.}
\label{positrons}
\end{figure*}
\begin{figure*}
 \centering
 \includegraphics[width=18cm,height=9cm,angle=0.0]{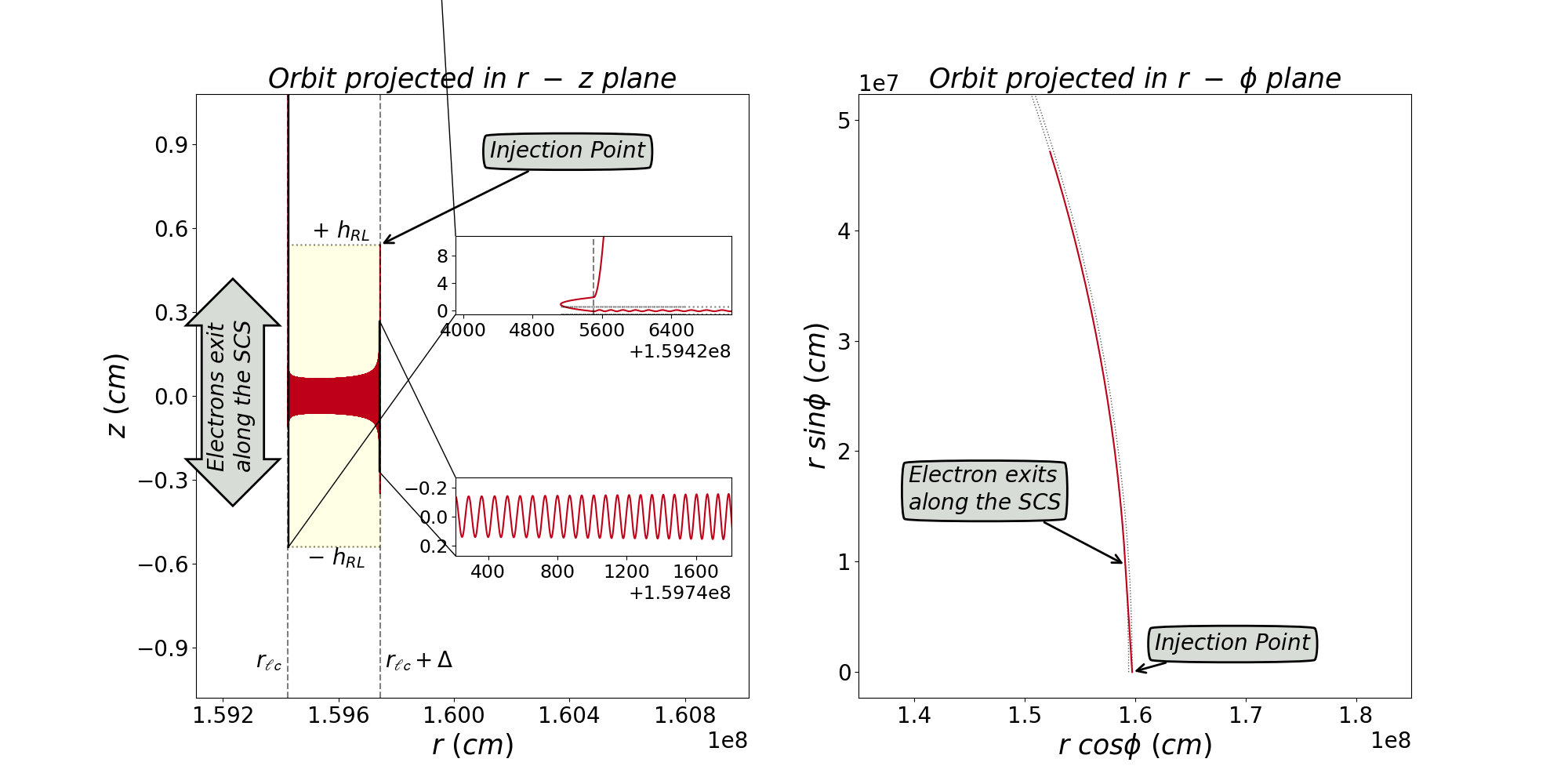}
\caption{Similar to figure~\ref{positrons} for electrons injected at the outer edge of the equatorial reconnection layer (red line). After they exit the reconnection layer, they enter the closed-line region (CLR) of strong magnetic field. They are then deflected in and out of the CLR and out of the equator along the boundary of the CLR where they support the negatively charged separatrix electric current-sheet (SCS), radiate away most of their energy, and return to the star.
}
\label{electrons}
\end{figure*}
\begin{figure*}
 \centering
 \includegraphics[width=18cm,height=9cm,angle=0.0]{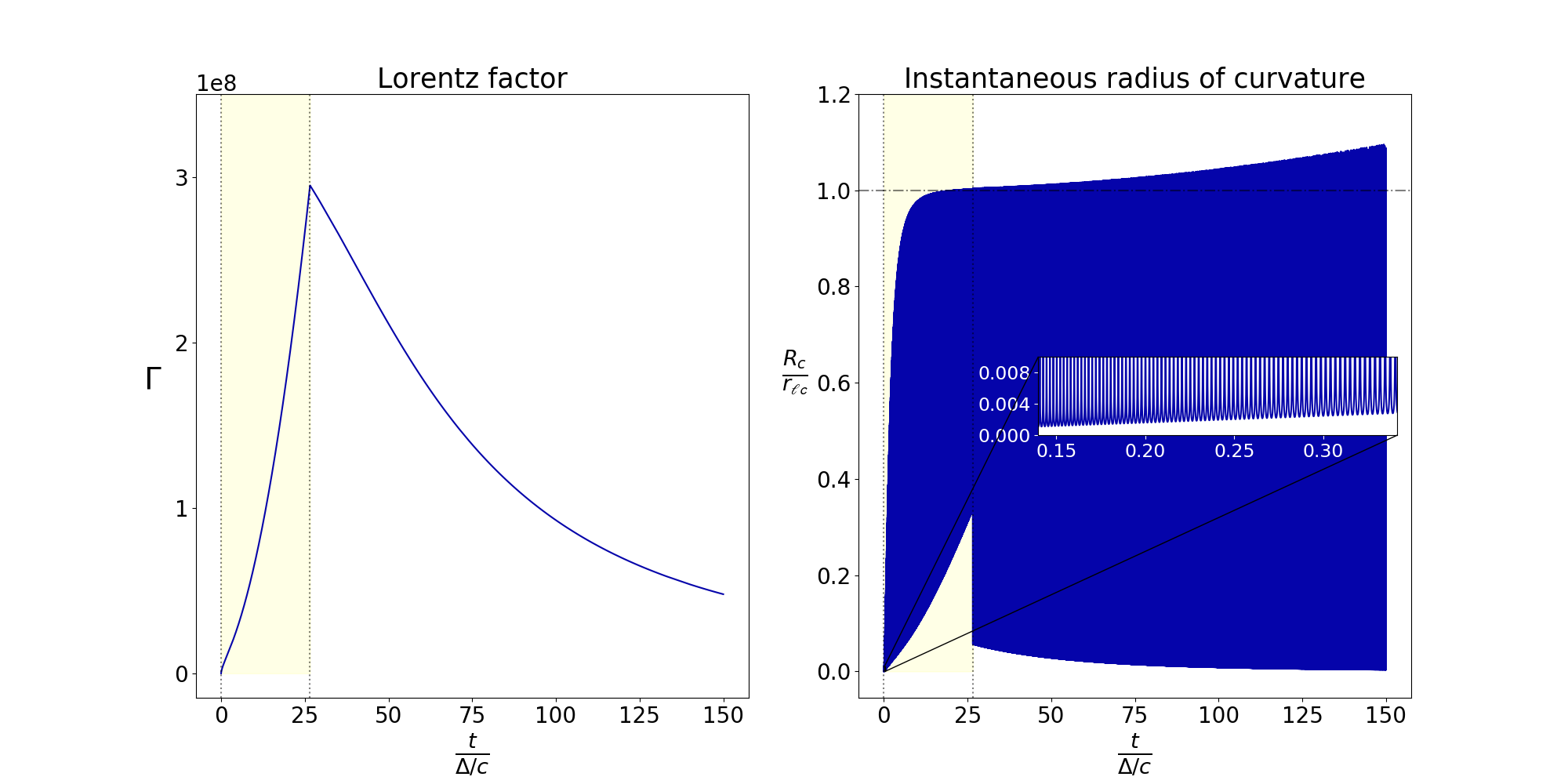}
\caption{Evolution of the Lorentz factor $\Gamma$ (left plot) and the instantaneous radius of curvature $R_{\rm c}$ (right plot) with time along the positron orbit of figure~\ref{positrons}. $R_{\rm c}$ in units of $r_{\rm lc}$. Time in units of the dissipation layer radial light-crossing time $\Delta/c$. Yellow time interval: time inside the dissipation layer. The detail corresponds to the detail of figure~\ref{positrons}.}
\label{GammaRcp}
\end{figure*}
\begin{figure*}
 \centering
 \includegraphics[width=18cm,height=9cm,angle=0.0]{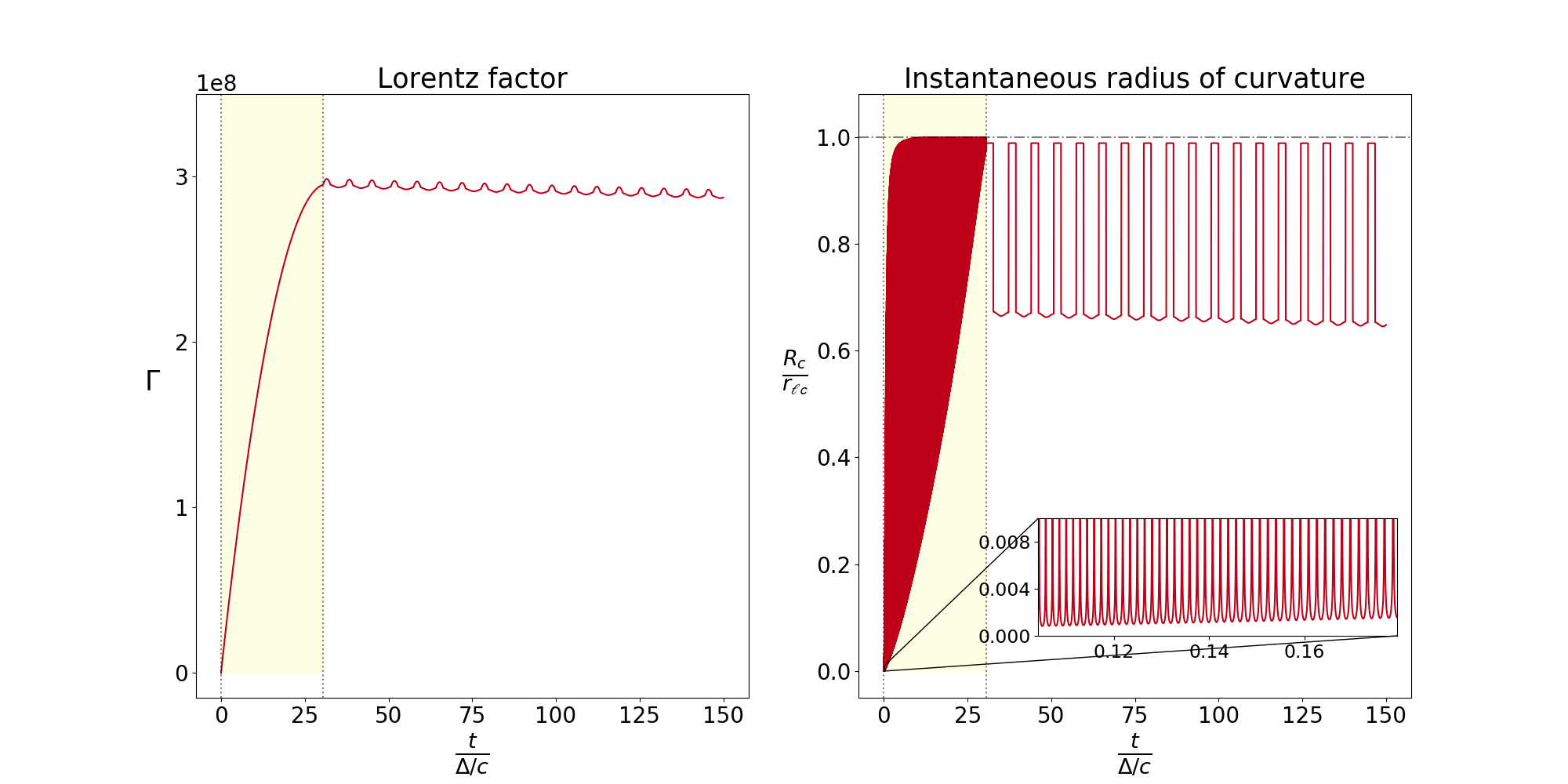}
\vspace{1cm}
\caption{Similar to figure~\ref{GammaRcp} for the electron orbit of figure~\ref{electrons}. The detail corresponds to the detail of figure~\ref{electrons} near the injection point of the orbit.}
\label{GammaRce}
\end{figure*}

\section{Particle trajectories}
\label{section}

It is straightforward to calculate the trajectories of electrons and positrons as they enter and exit the reconnection layer in the framework of our simple model of the ring-of-fire.
We will work in cylindrical coordinates $(r,\phi,z)$. According to the discussion in the previous section and Paper~I, magnetic and electric fields have different expressions in three equatorial regions very close to the light cylinder. In the interval $r_{\rm lc}< r < r_{\rm lc}+\Delta$ where lies the reconnection layer,
\begin{eqnarray}
B_\phi & = & -B_{\rm lc}\left(\frac{z}{h_{\rm RL}}\right)\ \mbox{when}\ z\leq |h_{\rm RL}| \nonumber\\
& = &  -B_{\rm lc}\mbox{sign}(z)\ \mbox{when}\ z>|h_{\rm RL}|\ ,
\label{2}\\
B_z & = & -B_{\rm lc}\ ,\\
E_r & = & -B_z\frac{r}{r_{\rm lc}}\ ,\\
B_r & = & E_\phi = E_z = 0\ ,
\end{eqnarray}
where, $B_{\rm lc} \equiv B_*(r_*/r_{\rm lc})^3/2$ is the dipole magnetic field at the distance of the light cylinder, $B_*$ is the magnetic field at the poles of the neutron star, and $h_{\rm RL}$ is the half-height of the reconnection layer (see below). In a slight departure from Paper~I, we assume here that the magnetic field crosses the reconnection layer perpendicularly.
$B_{\rm lc}$ is a typical value for $B_\phi$ and $B_z$ at the light cylinder. In reality, we expect that $B_\phi$ above the reconnection layer and $B_z$ will be different but close to each other. Their exact values can only be obtained via a detailed numerical simulation, but are not too important in our present work as long as they do not differ much from $B_{\rm lc}$.
The reconnection layer's radial electric current is supported by the gyration motion of the particles that enter it and move along it. Particles enter the reconnection layer with a distribution of initial Lorentz factors $\Gamma_{\rm inj}$ achieved at their point of origin in the polar cap acceleration gap \citep[e.g.][]{HMZ02}, and therefore, perform a distribution of gyration motions. 

In the present work, we assume that the half-height of the reconnection layer is equal to the gyroradius of the average injection Lorentz factor $\langle\Gamma_{\rm inj}\rangle$, namely
\begin{eqnarray}\label{h}
h_{\rm RL}=\frac{\langle\Gamma_{\rm inj}\rangle m_ec^2}{eB_{\rm lc}}\ .
\end{eqnarray}
Particles that enter the reconnection layer with $\Gamma_{\rm inj}\ll \langle\Gamma_{\rm inj}\rangle$ will remain much closer to the equator than particles that enter it with $\Gamma_{\rm inj}\gg \langle\Gamma_{\rm inj}\rangle$, and this is indeed observed in our calculations of particle trajectories. In a future work, we will calculate self-consistently the height and morphology of the reconnection layer by statistically adding all the particle orbits that enter it at all radii. This will yield a certain vertical distribution of the radial electric current density $J_r(z)\equiv(c/4\pi)({\rm d} B_\phi/{\rm d}z)$ and a corresponding vertical distribution of the azimuthal magnetic field accross the dissipation layer that will reverse sign on the midplane $z=0$, and will approach $B_{\rm lc}$ and $-B_{\rm lc}$ as $z\ll -h_{\rm RL}$ and $z\gg h_{\rm RL}$ respectively. Here, for simplicity, we have assumed a vertical azimuthal magnetic field distribution of the form of eq.~(\ref{2}).

Notice that we are not performing an evolutionary PIC simulation, since we are interested only in the acceleration and radiation of the particles. In the future, we hope to be able to perform self-consistent hybrid ideal force-free--PIC simulations in which the dynamical behavior of the dissipation zone will also be investigated. Our present approach differs from previous works in that, due to the vertical magnetic field  $B_z$ that threads the equatorial plane near the light cylinder, the reconnection layer is not Harris-type, and there is no region where $B=0$ and at the same time $E\neq 0$ as in most initial setups of reconnection layers without a guide field found in the literature \citep[e.g.][]{Cetal13}. The electric potential difference between the inner and outer radial boundaries of the reconnection layer is carried over from the ideal force-free region above and below the equator.

In the equatorial region near and beyond the reconnection layer $r\gsim r_{\rm lc}+\Delta$ lies the dissipationless current-sheet where
\begin{eqnarray}
B_\phi & = & -B_{\rm lc}\left(\frac{z}{h_{\rm ECS}}\right)\ \mbox{when}\ z\leq |h_{\rm ECS}| \nonumber\\
& = &  -B_{\rm lc}\mbox{sign}(z)\ \mbox{when}\ z>|h_{\rm ECS}|\ ,
\end{eqnarray}
and field lines open up to infinity, thus
\begin{eqnarray}
B_r = B_z = E_r = E_\phi = E_z = 0\ .
\label{8}
\end{eqnarray}
The half-height $h_{\rm ECS}$ of the equatorial current sheet is significantly smaller than $h_{\rm RL}$ due to the concentration of the positron orbits near the equator (see figure~\ref{positrons} below). Notice that $B_r$ and $E_z$ just above the equatorial current-sheet will become non-zero further away \citep[see e.g.][]{T06}. We acknowledge once again that the physical setup of the ring-of-fire and Paper~I (expressed mathematically by eqs.~\ref{2}--\ref{8}) is only approximate (see Appendix~B for a more detailed discussion). Nevertheless, we expect that the resulting particle acceleration and radiation will help us understand how the high-energy emission is produced in a real pulsar.

At the tip of the CLR $r\lsim r_{\rm lc}$ near and interior to the reconnection layer lies a region of compressed magnetic flux with
\begin{eqnarray}
|B_z| & \gg & B_{\rm lc}\ ,\label{discontinuity}\\
E_r & = & -B_z\frac{r}{r_{\rm lc}}\ ,
\end{eqnarray}
\begin{eqnarray}
B_r =B_\phi=E_\phi = E_z = 0\ .
\label{11}
\end{eqnarray}
The equatorial magnetic field abruptly increases by a very large factor as we move inwards accross the separatrix at the tip of the CLR near the light cylinder (see also Appendix~B). We arbitrarily take $|B_z|=10 B_{\rm lc}$ to simplify the calculation of electron trajectories below. This abrupt increase is a relativistic feature related to the presence of the separatrix electric current near the light cylinder where $|E_r|\rightarrow |B_z|$, and involves only a rearrangement of the magnetic flux at the tip of the CLR. This effect was predicted by \citet{U03}, but cannot be discerned numerically unless the numerical grid resolution is higher than a few thousand cells per $r_{\rm lc}$ (\citet[][fig.~1c]{S06} and \citet[][fig.~11]{T06}). 

We now proceed to calculate the trajectories of particles that enter and traverse the equatorial dissipation region. We will consider physical parameters relevant to the Crab pulsar, namely
\begin{eqnarray}
B_*=10^{13}\ {\rm G}\ ,\ P=0.033\ {\rm s}\ ,\ \kappa=500\ ,\  \mbox{and}\  \langle\Gamma_{\rm inj}\rangle=500\ .
\label{Crabparameters}
\end{eqnarray}
Here, $P\equiv2\pi/\Omega$ is the period of stellar rotation. $r_{\rm lc}=1.6\times 10^8$~cm, $\Delta=r_{\rm lc}/\kappa=3.2\times 10^5$~cm, $B_{\rm lc}=10^6$~G, $h_{\rm RL}\approx 1$~cm. Particles are injected just above the reconnection layer with a given initial Lorentz factor $\Gamma_{\rm inj}$. We assume that, by the time they reached that point during their travel from the polar cap to the ring-of-fire, they have lost to radiation their gyrating motion around the magnetic field. Therefore, at the injection point, the particle velocity $v$ consists of two components, one equal to the drift velocity
\begin{eqnarray}
&&v_{{\rm d}\ {\rm inj}} = c\frac{|E_r B_z|}{B_{\phi}^2+B_z^2}
\end{eqnarray}
perpendicular to the magnetic field in the $(\phi, z)$ plane, and one along the magnetic field, 
\begin{eqnarray}
&&v_{||\ {\rm inj}} = c\left(1-\Gamma_{\rm inj}^{-2}-\frac{E_r^2}{B_{\phi}^2+B_z^2}\right)^{1/2}\ .
\end{eqnarray}
We calculated $v_{||\ {\rm inj}}$ through the relation $v_{||\ {\rm inj}}^2+v_{{\rm d}\ {\rm inj}}^2=c^2(1-\Gamma_{\rm inj}^{-2})$. The velocity components at the injection point are
\begin{eqnarray}
v_{r\ {\rm inj}} & = & 0\ ,\\
v_{\phi\ {\rm inj}} & = & \frac{v_{||\ {\rm inj}}|B_\phi|-v_{{\rm d}\ {\rm inj}}|B_z|}{(B_\phi^2+B_z^2)^{1/2}}\ ,\\
v_{z\ {\rm inj}} & = & \frac{-v_{||\ {\rm inj}}|B_z|-v_{{\rm d}\ {\rm inj}}|B_\phi|}{(B_\phi^2+B_z^2)^{1/2}}\ .
\end{eqnarray}
Beyond the injection point we integrate the three components of the momentum equation for electrons and positrons with radiation reaction, namely 
\begin{eqnarray}\label{orbit}
m_e\frac{{\rm d}{\bf u}}{{\rm d} t}=\pm e\left( {\bf E}
+\frac{{\bf u}\times {\bf B}}{\Gamma c}\right)-\frac{P_{\rm rad}{\bf u}}{\Gamma c^2}\ ,
\end{eqnarray}
where, ${\bf u}\equiv \Gamma {\bf v}$ are the spatial components of the particle four velocity. The sign $+/-$ corresponds to positrons/electrons respectively, and $P_{\rm rad}$ is the power radiated by the accelerated particles,
\begin{eqnarray}
&&P_{\rm rad} \equiv \frac{2e^2 c \Gamma^4}{3 R_{\rm c}^2}\ ,
\label{Prad}
\end{eqnarray}
where the radius of curvature of the particle orbit is equal to
\begin{eqnarray}
&&R_{\rm c}
=\frac{c^3}{|{\bf v}\times ({\rm d}{\bf v}/{\rm d}t)|}
\label{Rc}
\end{eqnarray}
This general expression does not distinguish between synchrotron and curvature radiation.

Inside the reconnection layer, positrons are accelerated outwards and electrons are accelerated inwards. In figure~\ref{positrons}, we plot the 3D orbit of a positron entering the reconnection layer at its inner radius on the light cylinder with $\Gamma_{\rm inj}=500$. Similarly, in figure~\ref{electrons}, we plot the 3D orbit of an electron entering the reconnection layer at its outer radius with the same injection Lorentz factor. Inside the reconnection layer, both types of particles follow so-called Speiser orbits \citep{S65,C07b,Cetal13}, i.e. gyrating orbits that become more and more stretched and compressed towards the equatorial plane where the only remaining field components are the vertical magnetic field $B_z$, and the accelerating electric field $E_r$. Notice how small is the height compared to the length of these trajectories in figures~\ref{positrons} and \ref{electrons}. Notice also the overall deflection of the orbits in the azimuthal direction due to the overall pulsar rotation and the cylindrical nature of the problem.

After the positrons exit the reconnection layer at its outer radius, they form the positively charged dissipationless equatorial current-sheet where they gradually radiate away their remaining gyration motion, but experience no further acceleration nor orbital deflection. Notice that the half-height of the equatorial current sheet is much smaller than the half-height of the reconnection layer due to the vertical compression of the Speiser orbits towards the equator in the reconnection layer. We took $h_{\rm ECS}=0.1h_{\rm RL}$.

After the electrons exit the reconnection layer at its inner radius, they enter the tip of the CLR where they move along the strong vertical magnetic field and at the same time are quickly deflected back outwards by it. When they exit the CLR, the are deflected away from the equatorial plane by the azimuthal component of the magnetic field $B_\phi$. As a result of the two deflections, they form the negatively charged separatrix electric current-sheet and return to the star, gradually radiating away their remaining gyration motion. Obviously, in a real pulsar, the abrupt transitions shown in figures~(\ref{positrons}--\ref{electrons}) 
will be smoothed out.

Before we proceed with the calculation of the high-energy emission, we would like to argue why it is natural and of paramount importance that the acceleration zone lies outside the light cylinder. If the acceleration layer were inside the light cylinder, the accelerating electric field $E_r=-(r/r_{\rm lc})B_z$ would be smaller (in magnitude) than $B_z$, and because the overall particle orbits are deflected in the $\phi$ direction, very quickly a situation would be reached such that the centrifugal force term $+m_e \Gamma v_\phi^2/r$ balances the radial electromagnetic force $\pm e(E_r+v_\phi B_z/c)=\pm (v_\phi-r\Omega)e B_z/c$ in eq.~(\ref{orbit}). Most electrons and positrons would not be able to reach the ends of the acceleration zone where they are needed to support the separatrix and equatorial electric currents respectively. Moreover, their acceleration would stop at Lorentz factors much lower than the maximum one they could have reached had they moved all the way accross the acceleration zone. If the acceleration zone lies just outside the light cylinder as proposed in Paper~I, both problems are solved (see, however, also footnote~\ref{footnote} in Appendix~B). The electrons and positrons that enter the acceleration zone can fulfill their purpose, namely to support the magnetospheric electric current-sheets, and at the same time benefit from the full electric potential drop accross the acceleration zone.

\section{High-energy spectra}

The accelerated particles follow curved gyrating trajectories and emit high-energy radiation along the instantaneous direction of their motion. The radiation spectrum emitted at every point of the orbit depends on the instantaneous Lorentz factor $\Gamma$. In figure~\ref{GammaRcp}, we plot the evolution with time of the Lorentz factor and radius of curvature of the positron injected at the inner radius of the reconnection layer shown in figure~\ref{positrons}. Similarly, in figure~\ref{GammaRce} we plot the same for the electron injected at the outer radius of the reconnection layer shown in figure~\ref{electrons}. The details in figures~\ref{GammaRcp} and \ref{GammaRce} correspond to the details in figures~\ref{positrons} and \ref{electrons} respectively. We see that the radii of curvature of the Speiser orbits inside the reconnection layer nowhere exceed $r_{\rm lc}$ (this is due to the overall curvature of the orbit in the $\phi$ direction), whereas the minimum radii of curvature are initially one order of magnitude smaller, but they too reach values close to $r_{\rm lc}$ near the exit of the reconnection layer. Outside the reconnection layer, the positrons quickly lose their energy with radii of curvature $R_{\rm c}\ll r_{\rm lc}$. Some authors propose to characterize radiation that originates in parts of the orbit with smaller/larger $R_{\rm c}$ as synchrotron/curvature respectively \citep[e.g.][]{PS18}. 
\begin{figure}
 \centering
 \includegraphics[width=9cm,height=7.5cm,angle=0.0]{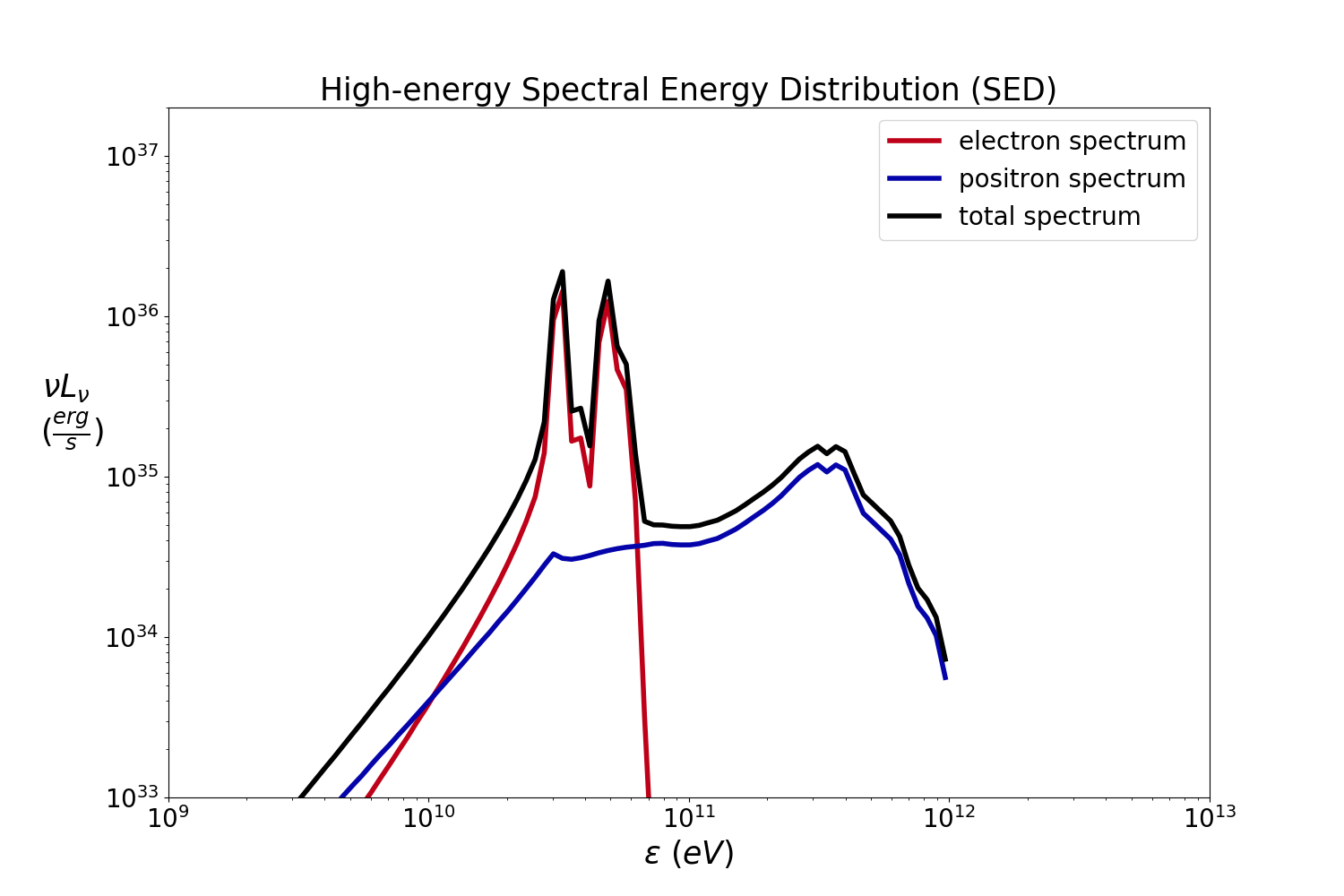}
\caption{Calculated high-energy Spectral Energy Distribution (SED) $\nu L_\nu$ in erg~s$^{-1}$. Blue line: positron contribution. Red line: electron contribution. Black line: total SED.
The VHE component that extends to the TeV range is due to the positrons that radiate away most of their energy along the equatorial current sheet. The electrons must travel a large distance along the separatrix return current sheet before they radiate away most of their energy, possibly in a different part of the spectrum. 
}
\label{SED}
\end{figure}

We generate high-energy spectra as follows: We assume that during a time interval ${\rm d} t$, a particle radiates a parcel of energy equal to ${\rm d} \epsilon = {\rm d} t P_{\rm rad}$ at the instantaneous  cutoff energy
\begin{eqnarray}
&&\epsilon_{\rm cut}=\frac{3\Gamma^3 hc}{eR_{\rm c}}
\end{eqnarray}
along its instantaneous direction of motion. We calculate for a long enough integration time about one hundred particle orbits with a Maxwell distribution of $\Gamma_{\rm inj}$ (positrons/electrons are injected at the inner/outer boundary of the reconnection layer respectively). We add the energy contributions from all photons emitted during the time interval ${\rm d}t$ distributed in particular energy bins with corresponding energy widths\footnote{In the present work we do not distinguish between directions of emission, and add all the photons. We plan to generate phase-resolved spectra in a future work.}. We divide the total collected energy in each bin by the width of the bin, and rescale the number of radiated parcels of energy to the total number of electrons and positrons that enter the equatorial dissipation zone during the above time interval (namely $2\pi (2\kappa \Omega B_{\rm lc}/2\pi ce)r_{\rm lc}\Delta c{\rm d}t=2\kappa B_{\rm lc}\Delta c{\rm d}t/e$). We then multiply by $\epsilon$, divide by ${\rm d}t$ and thus obtain the Spectral Energy Distribution (SED) of the total emitted high-energy radiation. In figure~\ref{SED} we plot the SED that we obtained for the particular Crab-like parameters in eqs.~(\ref{Crabparameters}) in ergs per second. The VHE component up to about 1~TeV is produced by the positrons that are accelerated outwards in the reconnection layer, and subsequently lose most of their energy along the equatorial current sheet as shown in figures~\ref{positrons} and \ref{GammaRcp}.

We acknowledge an important limitation of our analysis, namely that we are not allowed to integrate particle orbits very far from the ring-of-fire where the electric and magnetic fields begin to diverge from the simple expressions of eqs.~(\ref{2}--\ref{11}).
More precisely, within the calculated part of their trajectory, the outward moving positrons managed to radiate away the largest part of the energy they gained in the reconnection layer, and their contribution to the high-energy SED is shown in figure~\ref{SED}. On the contrary, within the same integration time, the electrons radiated away only a very small fraction of their energy. This suggests that they will need to travel a large distance along the separatrix return current sheet before they radiate away most of their energy. The conditions in the SCS are very different from those in the ECS, and therefore, the electron contribution to the high-energy SED is expected to be very different from that of the positrons (it may, for example, account for the part of the SED that peaks around a few hundred keV). If we want to reproduce the detailed phase-resolved SED of a real pulsar, we must carefully take into account the detailed 3D geometry, the distribution of $B_{\rm lc}$ values around the ring-of-fire, and the photon direction and travel time. This will be the subject of a future work. 

\section{Summary}

In the present work we have proposed a simple model according to which most of the particle acceleration in pulsars takes place in a narrow zone beyond the tip of the corrotating closed-line region (CLR) where it touches the light cylinder, a so-called ring-of-fire. This is the main magnetospheric dissipation zone inside the pulsar wind termination shock. Its radial width is inversely proportional to the pair-formation multiplicity $\kappa$ at the electrostatic gap above the polar cap.  $\kappa$ also determines the dissipation efficiency, thus also the efficiency $\eta_\gamma$ for gamma-ray production in pulsars ($\eta_\gamma \sim 1/2\kappa$; see Paper~I). We have shown that most of the observed high-energy radiation originates in two extended regions adjacent to the particle acceleration zone, one outside where the outwards accelerated positrons radiate in the equatorial current-sheet, and one just inside the light cylinder where the inwards accelerated electrons radiate along the separatrix.

In our model, we have essentially separated the regions where particles emit high-energy radiation from the narrow region where they are accelerated by the pulsar electromagnetic energy-loss. In reality, the acceleration region will be more extended, the two processes (acceleration and emission of radiation) will take place at the same time, and the SED will be smoother than the one shown in figure~\ref{SED}. Obviously, our model cannot substitute the effort invested in numerical pulsar calculations. Our goal was not to obtain a phenomenological reproduction of the high-energy spectrum of a particular pulsar \citep[as e.g. in][]{T18}, but to attain a deeper physical understanding of the magnetospheric conditions around the tip of the CLR near the light cylinder. We learned a lot of things with our model, namely what are the main regions of particle acceleration and radiation, what are the maximum Lorentz factors and $\gamma$-ray energies that can be attained ($\sim 10^8$ and $\sim 1$~TeV respectively), what parts of the spectrum are generated by each particle type, etc. Our next target is to obtain more realistic pulsar spectra by investigating the structure of the ring-of-fire in greater detail, both with semi-analytic calculations (as the ones presented here) and self-consistent hybrid ideal force-free--PIC simulations.

\section*{Acknowledgements}

We acknowledge interesting discussions with Drs. Constantinos Kalapotharakos, Demosthenes Kazanas, Apostolos Mastichiadis, and Nektarios Vlahakis. 


\bibliographystyle{mn2e}

\section*{Appendix~A: Global energetics of reconnection layers} 

We would like to point out here two important issues that we believe are not dealt with adequately in local numerical investigations of reconnecting layers. For simplicity, we will only consider axisymmetric configurations with radial equatorial current-sheets. Our conclusions can be directly generalized in general current-sheets.

The first issue has to do with the total amount of charge carriers available in the plasma that are transported into the current-sheet, and the amount of charge carriers needed to support the electric current in the current-sheet. In a reconnecting layer, all charge carriers are transported into the current-sheet at a rate of
\begin{equation}\label{rate}
2nc\frac{E_r B_\phi}{B^2}\ \mbox{charge carriers per unit surface.}
\end{equation}
Here, $n$ is the total number density of charge carriers in the plasma, and the factor of 2 takes into account the contributions from both surfaces of the current-sheet. If the reconnection region has a total radial extent $\Delta$, and if electric charge carriers are provided {\it only in that region}, then the equatorial electric current is equal to
\begin{equation}\label{dr}
\frac{1}{2}\frac{2enc E_r B_\phi}{B^2}2\pi r\Delta = I =
crB_\phi\ .
\end{equation}
The factor of $1/2$ is there because the charge carriers consist of an (almost) equal number of positive and negative charges. Each type exits from the two opposite sides of the current-sheet in the $r$-direction, and both types contribute to the same radial electric current $I$. Eq.~(\ref{dr}) poses a very strong restriction on the radial extent of the current-sheet, namely that
\begin{equation}\label{deltar}
n\Delta = \frac{B^2}{2\pi e E_r}\ .
\end{equation}
In the present case of pulsar current closure, 
\[
E_r= B_z\left(\frac{r}{r_{\rm lc}}\right)\approx B_z\ ,
\]
\[
B^2=B_z^2+B_\phi^2\approx 2B_z^2\ ,
\]
\[
n=2\kappa n_{\rm GJ}=\frac{\kappa\Omega  B_z}{\pi ec}\ ,\label{n}
\]
and therefore,
\begin{equation}\label{Delta}
\Delta \approx \frac{c}{\kappa \Omega}=\frac{r_{\rm lc}}{\kappa}
\end{equation}
(see eq.~15 in Paper~I). Therefore, our choice of the radial extent of the reconnecting layer satisfies the condition that the total amount of charge carriers that are transported into the current-sheet is precisely equal to the amount of charge carriers needed to support its electric current. This was one of the main points of Paper~I. In previous numerical simulations of reconnection layers in the literature, no restriction is imposed between the density of charge carriers in the reconnecting plasma and the length of the reconnection layer in the direction of its electric current, as e.g. in plane-parallel local numerical studies where this length is mathematically infinite. In fact, we are wondering whether certain features of local numerical simulations, in particular the formation and conglomeration of plasmoids, may have anything to do with the availability of more charge carriers than the precise number needed to support its electric current as in eq.~(\ref{deltar}) above. It is reasonable to expect that the extra pairs of oppositely charged particles (electrons and positrons, or electrons and ions) generate local electric currents that are in excess of the overall electric current in the current-sheet (eq.~\ref{dr} in the present configuration), thus forming plasmoids that contain the extra current and keep growing as more and more excessive pairs of oppositely charged particles are accumulated there. Equivalently, a reconnecting layer with radial extent as specified in Paper~I and eq.~(\ref{Delta}) will not break down into plasmoids. This issue is worthy of further investigation.

The second issue has to do with the electromagnetic energy flux that enters the current-sheet and must be dissipated in the acceleration of particles. If the current-sheet has a radial extend $\Delta$, the total amount of Poynting flux that enters it is equal to
\begin{equation}
2c \frac{E_r B_\phi}{4\pi} 2\pi r \Delta\ .
\end{equation}
Let us denote here by $\epsilon$ the average amount of energy that every particle that enters the current-sheet absorbs on average. We need to perform detailed trajectory calculations to determine how much energy is gained by particles entering the current-sheets at different radial positions. Nevertheless, it is easy to understand that the maximum available energy for a particle to attain is $eE_r\Delta$, but particles enter the current layer uniformly at all radial positions, thus the average energy each one attains is only $1/2$ of the maximum available. Therefore, $\epsilon=eE_r\Delta/2$. We must then satisfy that
\begin{equation}\label{Poynting}
cE_r B_\phi r \Delta = \frac{2\epsilon ncE_r B_\phi}{B^2} 2\pi r \Delta=\frac{2\pi encrE_r^2 B_\phi \Delta^2}{B^2}\ ,
\end{equation}
from which we recover eq.~(\ref{deltar}). We have thus shown that the issue of precise charge carrier availability required to form the electric current is equivalent to the global 
energetics of the reconnection layer. In other words, the energy needed to accelerate all the particles that enter the current sheet to one half its maximum accelerating potential is precisely equal to the electromagnetic (Poynting) energy that enters it from above and below.

\section*{Appendix~B: The boundaries of the reconnection layer}

In the ideal force-free steady-state axisymmetric solution of \citet{CKF99}, the CLR extends all the way to the light cylinder, and lies adjacent to the region of open magnetic field lines. Between the two develops the return-current separatrix accross which, we must guarantee continuity of $B^2-E^2$. \citet{U03} investigated this particular problem and concluded that a magnetic field discontinuity develops accross the separatrix in the equator, namely that $B_z$ vanishes just outside but diverges as $1/\sqrt{1-(r/r_{\rm lc})^2}$ just inside the separatrix. In the present work and in Paper~I, the CLR also extends all the way to the light cylinder, but adjacent to it now lies a narrow region of magnetic field lines that, instead of extending outwards to infinity, cross the equator in a narrow dissipation zone just outside the light cylinder. Beyond that region magnetic field lines open up to infinity. Between that region and the CLR again develops the return-current separatrix. As before, we must guarantee continuity of $B^2-E^2$ across the inner and outer boundaries of the dissipation zone.

We can directly satisfy continuity of $B^2-E^2$ at the inner boundary. Just inside it, at $r=r_{\rm lc}-\delta$ ($\delta\ll r_{\rm lc}$), $B^2-E^2=B_z^2-E_r^2=B_z^2(1-[r/r_{\rm lc}]^2)\approx (2\delta/r_{\rm lc})B_z^2$. For a sufficiently large discontinuity of $B_z$ accross the width of the separatrix return-current-sheet, this can be made equal to $B^2-E^2\approx B_\phi^2$ just outside the light cylinder (eq.~\ref{discontinuity}; see \citet{U03} for details). There is, however, a small problem at the outer boundary beyond the light cylinder at $r=r_{\rm lc}+\Delta$. Just inside it, $B^2-E^2=B_{\rm lc}^2(1-[r/r_{\rm lc}]^2)+B_\phi^2\lsim B_\phi^2$, whereas just outside it, $B^2-E^2=B_\phi^2$.

In realistic numerical simulations\footnote{\label{footnote}
We would like to discuss here another secondary feature of such simulations, namely that the inner edge of the dissipation zone lies consistently a certain distance {\it inside} the light cylinder. This is most probably due to the inwards accelerated electrons reaching the centrifugal barrier discussed above at the end of section~\ref{section} (this is clearly seen at the point denoted with a small dark triangle pointing to the left in the left panels in Fig.~14 of \citet{CPPS15}, and at the point denoted with a small dark rhombus in the right panels of the same figure). Obviously, in our case too, the inner edge of the dissipation layer may extend some distance inside the light cylinder. We found, however that, for realistic Crab-like pulsar parameters, the radial  distance needed for the inwards accelerated electrons to reach the centrifugal barrier is much smaller than $\Delta$ (which is already much smaller than $r_{\rm lc}$). Therefore, the dissipation layer may indeed start some distance inside the light cylinder, but for realistic Crab-like pulsar parameters, that distance is expected to be very-very small.
}
\citep[e.g. Figs.~4 of][]{CPPS15}, this small discontinuity is amended with a horizontal tilt of the magnetic field. This provides a continuous poloidal magnetic field component on both sides of the last magnetic field line that enters the dissipation layer. Poloidal magnetic field lines (i.e. the $(B_r, B_z)$ component of ${\bf B}$) enter the equatorial dissipation zone perpendicularly at its inner edge just outside the tip of the CLR, and gradually enter it more and more horizontally as we approach its outer edge. In fact, it is conceivable that the dissipation zone extends from the light cylinder to infinity. However, every small amount of poloidal magnetic flux that ends up in the dissipation zone carries about the same amount of Poynting flux. Therefore, the largest fraction of the dissipation and the resulting particle acceleration is expected to take place within a distance of about $\Delta$ from the light cylinder where poloidal magnetic field lines enter the dissipation zone almost perpendicularly. Beyond that distance, there is only a small amount of extra dissipation and particle acceleration, both distributed gradually over a radial distance much larger than $\Delta$.

We, thus, conclude that our picture of a limited narrow dissipation region (where most of the particle acceleration takes place) and a dissipationless equatorial current-sheet outside (where the accelerated particles gradually lose most of their energy) is a simplification. Nevertheless, we expect that the more realistic setup described above will yield results (high-energy radiation spectra) similar to the ones obtained with our present simple model. We envision to investigate this problem further with future self-consistent hybrid force-free--PIC numerical simulations. 

\end{document}